# Lattice distortion effects on the frustrated spin-1 triangular-antiferromagnet $A_3NiNb_2O_9$ (A = Ba, Sr and Ca)

Z. Lu [1], L. Ge [2], G. Wang [3], M. Russina [1], G. Guenther [1], C. R. dela Cruz [4], R. Sinclair [5], H. D. Zhou [5], J. Ma [3,6]∗

1 Helmholtz-Zentrum Berlin für Materialien und Energie, D-14109 Berlin, Germany
2 School of Physics, Georgia Institute of Technology, Atlanta, GA 30332, USA
3 Key Laboratory of Artificial Structures and Quantum Control, School of Physics and Astronomy, Shanghai Jiao Tong University, Shanghai 200240, China
4 Neutron Scattering Division, Oak Ridge National Laboratory, Oak Ridge, Tennessee 37831, USA
5 Department of Physics and Astronomy, University of Tennessee, Knoxville, Tennessee 37996, USA
6 Collaborative Innovation Center of Advanced Microstructures, Nanjing, Jiangsu 210093, China

**Abstract**

In the geometrically frustrated materials with the low-dimensional and small spin moment, the quantum fluctuation could interfere with the complicated interplay of the spin, electron, lattice and orbital interactions, and host exotic ground states such as nematic spin-state and chiral liquid phase. While the quantum phases of the one-dimensional chain and S - ½ two-dimensional triangular-lattice antiferromagnet (TLAF) had been more thoroughly investigated by both theorists and experimentalists, the work on S = 1 TLAF has been limited. We induced the lattice distortion into the TLAFs, $A_3NiNb_2O_9$ ($A$ = Ba, Sr, and Ca) with $S$ ($Ni^{2+}$) = 1, and applied the thermodynamic, magnetic and neutron scattering measurements. Although $A_3NiNb_2O_9$ kept the non-collinear 120° antiferromagnetic phase as the ground state, the $Ni^{2+}$-lattice changed from the equilateral triangle ($A$ = Ba) into isosceles triangle ($A$ = Sr and Ca). The inelastic neutron scattering data were simulated by the linear spin-wave theory, and the competition between the single-ion anisotropy and the exchange anisotropy from the distorted lattice was discussed.

## I. INTRODUCTION

In the geometrically frustrated system, the complicated interaction(s) between electron, phonon, spin and orbital could lead to degenerate ground states, which introduce exotic properties and attracted a lot attention over the past decades [1-3]. Meanwhile, those degenerate states would easily be held by the symmetrical inconsistency and be destroyed by significant quantum fluctuations, not only induced by the complicated interactions among low dimensionality, geometrical frustration, small spin and the applied magnetic field, but also modified the classical Heisenberg model [4-5]. The triangular-lattice antiferromagnet (TLAF), one of the simplest frustrated two-dimensional (2D) material, had been suggested to be strongly influenced by the strong quantum spin fluctuations with small effective spin ($S = 1/2$ or 1) and exhibited a rich variety of interesting physics [6-8]. A striking example of these quantum phenomena is the transition from a non-collinear 120° spin configuration at 0 T into fractional-magnet lattice under a finite range of applied magnetic field, such as a collinear up-up-down (*uud*) phase corresponding to a magnetization plateau with one-third of its saturation value [9-13].

Recently, the theoretical research indicated the *uud* state as a commensurate analogue of the incommensurate spin-density-wave which was predicted and observed for frustrated one-dimensional spin-1 chains and $S = ½$ TLAF [14, 15], and suggested the possibility for exotic magnetic excitations. While there was an enormous theoretical activity in the domain, a full consensus on the origin of the *uud* state, (even the ground magnetic state, non-collinear 120º at zero field) was limited. This is true as well for the state-of-the art experimental investigation of its spin excitations due to the lack of a triangular-lattice materials. Although the lattice distortion has been believed to influence the quantum effects in the systems of kagome, square, and triangular motifs via the antisymmetric Dzyaloshinsky-Moriya (DM) interaction and requires that the dynamic models include the item of lattice contribution [16-20], how the ground state of triangular lattice originate from the lattice is still unclear. If the distortion effect was gradually introduced in the system, it would be insightful to obtain the specific physical properties as the related lattice is changed.

Ba$_3$NiNb$_2$O$_9$ was one of the first-studied equilateral TLAFs with highly symmetric crystal structure, which was free from the antisymmetric DM interaction such that a simple Hamiltonian model was expected to describe the physics of this material. The crystal structure is trigonal, P-3m1: i) the corner oxygen was shared by the NiO$_6$ and NbO$_6$ octahedra; ii) Ni ions occupied the 1$b$ Wyckoff sites and form the triangular lattice in the $ab$-plane; iii) Ba$^{2+}$ ions built up the unit cell frame and split NiO$_6$ octahedra. As the $ab$-planes were separated by double nonmagnetic layers consisting of the Nb$_2$O$_{11}$ double octahedra and Ba$^{2+}$ ions, it was expected that the inter-layer exchange interaction was much smaller compared to the nearest-neighbor (NN) exchange interaction in the $ab$-plane of Ni$^{2+}$ ions and the compound could be treated or approximately treated as a 2D system.

In this paper, we focused on the ground state of the non-collinear 120° spin structure reported for the triangular-lattice antiferromagnets $A_3$NiNb$_2$O$_9$ ($A$ = Ba, Sr, and Ca). Since the magnetic exchange energies were sensitive to the bond lengths/angles of Ni and O ions, it was interesting to compare the family of $A_3$NiNb$_2$O$_9$ ($A$ = Ba, Sr, and Ca) as the structures are gradually distorted. As Sr$^{2+}$ and Ca$^{2+}$ ions were smaller than Ba$^{2+}$ ion, the effect from the lattice distortion would be introduced gradually and the $A$-site effect on the exchange interactions were observed. In addition, a strong coupling between the successive magnetic phase transitions and the ferroelectricity had been observed in Ba$_3$NiNb$_2$O$_9$ and Sr$_3$NiNb$_2$O$_9$ by magnetic and electric bulk measurements [10, 21]. Our study on the lattice effect on the triangular lattice should not only discuss the quantum effect in TLAFs, but also be beneficial for the development of multiferroicity theory in low-dimensional frustrated materials [22].

## II. EXPERIMENTAL

Polycrystalline $A_3$NiNb$_2$O$_9$ ($A$ = Ba, Sr, and Ca) samples were prepared by solid state reaction method. Stoichiometric mixtures of BaCO$_3$/SrCO$_3$/CaCO$_3$, NiO and Nb$_2$O$_5$ were ground together, and calcined in air at 1230 ºC for 24 h. A commercial SQUID magnetometer (MPMS, Quantum Design) and a high-field vibrating sample magnetometer (VSM) were employed to measure the dc magnetization as a function of temperature and magnetic field. The specific heat was measured by the physical property measurement system (PPMS, Quantum Design) by two steps. First, the background specific heat was measured by an empty pucker with $N$-grease; then, a dense and solid thin plate of $A_3NiNb_2O_9$ sample with total mass around 10 mg was measured. After that, by subtracting the background specific heat from this total specific heat, we obtained the specific heat of the sample.

The neutron powder diffraction (NPD) measurements down to 0.3 K were performed using the HB-2A powder diffractometer at the High Flux Isotope Reactor (HFIR), Oak Ridge National Laboratory (ORNL), with a $^3$He insert system. About 5 g of pelletized powder for each sample was loaded in a vanadium can with He exchange gas. Data were collected at selected temperatures using two different wavelengths $\lambda$ = 1.538 and 2.406 Å. The collimation was set as open-open-6'. The shorter wavelength was used to investigate the crystal structures with the higher $Q$ coverage, while the longer wavelength was important for investigating the magnetic structures of the material with the lower $Q$ coverage. The Rietveld refinements on the diffraction data were performed using the program FullProf [23].

The inelastic neutron scattering (INS) of polycrystalline $A_3NiNb_2O_9$ ($A$ = Ba, Sr, and Ca) samples were carried out on recently upgraded cold neutron direct-geometry time-of-flight spectrometer NEAT at Helmholtz-Zentrum Berlin [24]. The spectrometer allows to cover a wide range of energy transfers $\hbar\omega$ and scattering angles, thereby allowing determination of a large swath of the scattering intensity $S(Q,\omega)$ as a function of momentum transfer $\hbar Q$ and energy transfer $\hbar\omega$, where $Q$ is the scattering vector. Around 4 g of each sample was packed in aluminum cans filled with He exchange gas. Each scan was counted around 6 hours with the incident neutron energy $E_i$ = 3.272 meV.

### III. RESULTS

**A. Neutron powder diffraction**

As shown in Figs. 1(a) and (b), low temperature NPD were employed at 8 and 0.3 K to identify the magnetic structure of polycrystalline $Ca_3NiNb_2O_9$. The space group could be indexed using a monoclinic unit cell, space group P 1 $2_1$/c 1 (No. 14), with one Ni atom at the $2a$ (0, 0, 0) site and another one at the $2d$ site (1/2, 1/2, 1/2) and the other Ca, Nb and

O atoms at the 8$f$ (x, y, z) site. By comparing the Rietveld refinement results of Ca$_3$NiNb$_2$O$_9$ patterns at 0.3 and 20 K (20 K data was not presented here), it suggested that no crystal structure (P 1 2$_1$/c 1) transition was observed down to 0.3 K. The (1/3, 1/3, 1/2) and (4/3, 1/3, 1/2) magnetic Bragg peaks were observed at 0.3 K and absent at 8 K with $Q$~0.85 Å$^{-1}$ and 2.0 Å$^{-1}$, which were consistent with the commensurate magnetic wave vector $q_m$= [1/3, 1/3, 1/2] (Fig. 1(b)), suggesting a similar 120° spin structure to Ba$_3$NiNb$_2$O$_9$ and Sr$_3$NiNb$_2$O$_9$ at 0.3 K [10, 18]. The magnetic phase transition was displayed by the temperature dependence of the order parameter (OP), Fig. 1(c) measured as the temperature dependence of the intensity of the [1/3 1/3 1/2] magnetic peak.

TABLE I. The crystallographic properties of $A_3$NiNb$_2$O$_9$ ($A$ = Ba, Sr, and Ca).

| Parameters | Ba$_3$NiNb$_2$O$_9$ | Sr$_3$NiNb$_2$O$_9$ | Ca$_3$NiNb$_2$O$_9$ |
|---|---|---|---|
| Space group | P -3 2/m 1 | P 1 2$_1$/c 1 | P 1 2$_1$/c 1 |
| lattice parameters | $a$=$b$=5.7509(5) Å<br>$c$=7.0343(8) Å | $a$=9.7549(5) Å<br>$b$=5.6387(1) Å<br>$c$=16.9194(8) Å | $a$=9.5695(8) Å<br>$b$=5.4472(2) Å<br>$c$=16.7876(3) Å |
| Interaxial angles | $\alpha$=$\beta$=90°<br>$\gamma$=120° | $\alpha$=$\gamma$=90°<br>$\beta$=125.066(4)° | $\alpha$=$\gamma$=90°<br>$\beta$=125.718(3)° |
| $t$ | 1.031 | 0.972 | 0.938 |

Fig. 1(d) presents the schematic crystal structure of $A_3$NiNb$_2$O$_9$. For the triple-perovskite system $A_3$NiNb$_2$O$_9$, the Ni-triangular layers were split by the nonmagnetic corner shared Nb$_2$O$_{11}$ perovskite. As the substitution of the smaller ions into the Ba-site with Ca$^{2+}$, the lattice of Ca$_3$NiNb$_2$O$_9$ was distorted from hexagonal (P -3 2/m 1) to monoclinic (P 1 2$_1$/c 1) space group as Sr$_3$NiNb$_2$O$_9$. Therefore, the equilateral Ni-triangle changed to the isosceles triangle by replacing the Ba$^{2+}$ ions with Sr$^{2+}$ and Ca$^{2+}$ ions, Fig. 1(e). The Ba compound exhibited an equilateral Ni-triangle with bonds of 5.7509 Å. For the Sr compound, one longer bond of 5.6387 Å and two shorter bonds of 5.6339 Å were obtained; for the Ca compound, one shorter bond of 5.4470 Å and two longer bonds of 5.5056 Å were observed.

The tolerance factor, $t$, suggested by Goldschmidt had been employed to describe the stability of perovskite phases: the deviation of $t$ from $t=1$ could be applied to estimate the internal strain in perovskites and oxygen octahedral tilt due to the misfit of the A and B site ionic radii [25]. The definition of $t$ was given by,

$$t = \frac{R_A + R_O}{\sqrt{2}(\langle R_B \rangle + R_O)}$$

where $\langle R_B \rangle$ is the average ionic radii for the B site ions [26], and $R_A$ and $R_O$ are the A and O site ionic radius, respectively.

The crystallographic properties of $A_3NiNb_2O_9$ ($A$ = Ba, Sr, and Ca) were given in TABLE I. For $Ba_3NiNb_2O_9$, the tolerance factor was bigger than 1 while it was smaller than 1 for $Sr_3NiNb_2O_9$. However, the $t$ deviations from $t=1$ were similar with ~0.030, which suggests a similar distortion from the ideal perovskite for them. Therefore, the crystal structure distorted from hexagonal to monoclinic structure is expected to result to a smaller $t$. Moreover, the angles of the equilateral triangle in $Ba_3NiNb_2O_9$ were all 60°; for $Sr_3NiNb_2O_9$, they were 60.090°, 59.955° and 59.955°; and for $Ca_3NiNb_2O_9$, they were 58.93°, 60.535°, and 60.535°.

TABLE II. The bond lengths of $A_3NiNb_2O_9$ ($A$ = Ba, Sr, and Ca).

| Bonds | $Ba_3NiNb_2O_9$ | $Sr_3NiNb_2O_9$ | $Ca_3NiNb_2O_9$ |
|---|---|---|---|
| <$Ba_1$/$Sr_1$/$Ca_1$-O> | 2.8695(1) | 2.7829(4) | 2.7345(3) |
| <$Ba_2$/$Sr_2$/$Ca_2$-O> | 2.8850(9) | 2.8243(4) | 2.8689(2) |
| <$Sr_3$/$Ca_3$-O> |  | 2.8962(6) | 2.8560(9) |
| <$Ni_1$-O> |  | 1.9416(9) | 2.1165(9) |
| <$Ni_2$-O> | 2.0901(3) | 2.0449(5) | 2.0653(4) |
| <Ni-O> |  | 1.9933(2) | 2.0909(6) |
| <$Nb_1$-O> | 1.9416(7) | 2.0004(7) | 2.0005(5) |
| <$Nb_2$-O> | 2.0662(2) | 2.0731(1) | 2.0625(8) |
| <Nb-O> | 2.0039(5) | 2.0367(9) | 2.0315(6) |

TABLE II showed the bond lengths of $A_3NiNb_2O_9$ ($A$ = Ba, Sr, and Ca). Considering the bond-length, the six Ni-O bonds had the same lengths of 2.090 Å in $Ba_3NiNb_2O_9$. However, the smaller ionic radius of $Nb^{5+}$, 0.64 Å, compared to 0.69 Å for $Ni^{2+}$ in octahedral coordination resulted in breathing distortion (extension or contraction of the Nb-O bond lengths) of the $NbO_6$ octahedra with two different lengths of Nb-O bonds, 1.942 and 2.066 Å, respectively. In $Sr_3NiNb_2O_9$ and $Ca_3NiNb_2O_9$, A-site ion is too small to hold its occupied cub-octahedral site, and then a combination of the breathing and tilting distortions both occurred, which can reduce the volume of the interstice and thereby improve the structural stability. With decreasing the size of the A-site ion, for $Sr_3NiNb_2O_9$ the average bond lengths of A-O and B-O decreased, but for $Ca_3NiNb_2O_9$ with smallest A-site ion the <A-O> and <B-O> are bigger than those for $Sr_3NiNb_2O_9$ due to stronger lattice distortion.

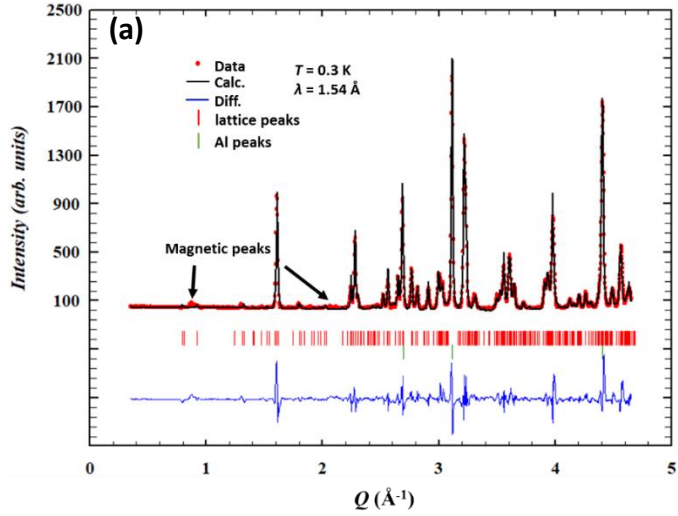

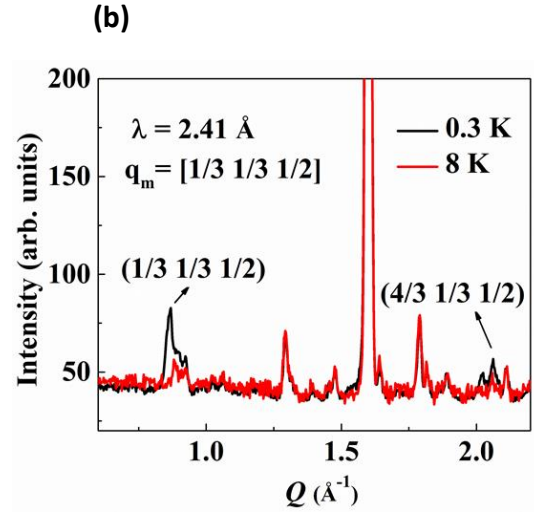

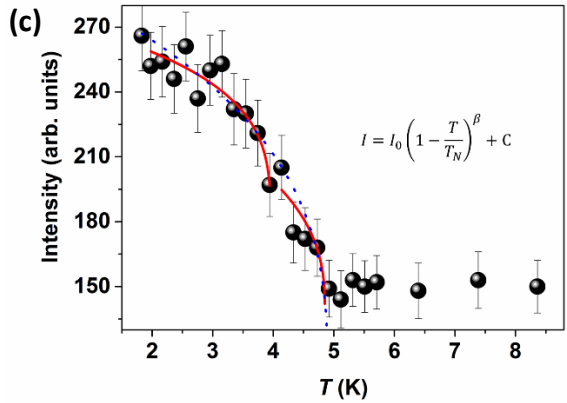

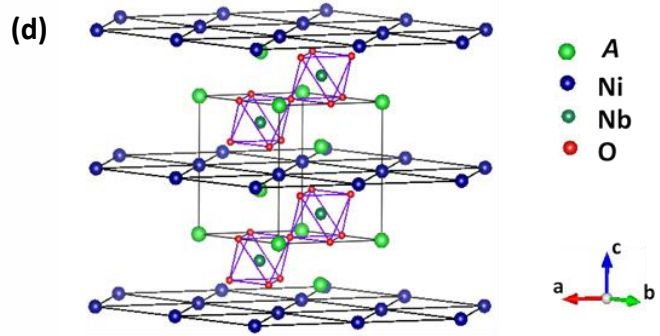

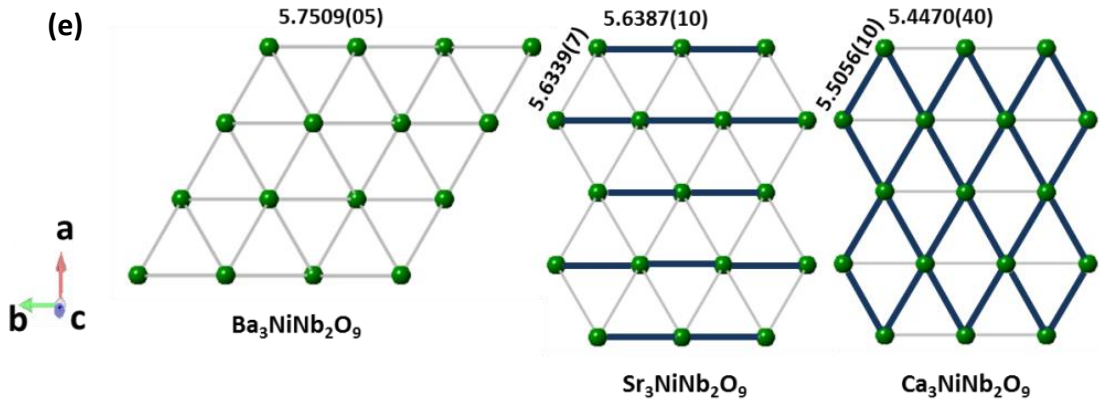

FIG. 1. (a) The neutron powder diffraction (NPD) pattern of polycrystalline $Ca_3NiNb_2O_9$ performed at 0.3 K. The refinement was done using the monoclinic space group, P 1 $2_1$/c 1. (b) The NPD patterns taken at 0.3 and 8 K. The (1/3, 1/3, 1/2) and (4/3, 1/3, 1/2) magnetic Bragg peaks at 0.3 K was given based on the hexagonal P-3m1 space group. (c) Temperature dependence of the order parameter in $Ca_3NiNb_2O_9$. Lines were fits to one (dashed) or two (solid) order parameters. The latter fit the data better suggesting a two-step transition consistent with what is observed in the bulk magnetization measurements discussed in the next section. (d) The schematic crystal structure of $A_3NiNb_2O_9$. (e) Triangle lattice of $Ni^{2+}$ ions in *ab*-plane of $A_3NiNb_2O_9$.

In Fig. 1(c), the fitting of OP was assumed with one or two order parameters, shown as dashed and solid lines, respectively. $T_{N1}$=4.90 K was obtained by the single-OP fitting with $\beta_{one}\approx 0.39$. The two-OP fitting gave $T_{N1}$=4.85 K and $T_{N2}$=4.05 K and yielded critical exponents, $\beta_{two-1}\approx 0.34$ and $\beta_{two-2}\approx 0.36$, respectively, which were a better description of the data. It was worth noting that the obtained values were comparable with those of 2-vector XY ($\beta$=0.35) or 3-vector Heisenberg ($\beta$=0.36) models. The latter fit the data better suggesting a two-step transition consistent with what is observed in the bulk magnetization measurements discussed in the next section. However, the effect of averaging in powder diffraction could only determine the 120° spin direction projected in the *ab*-plane and cannot determine whether the Ni-triangular layers are of the easy-axis or easy-plane type [27].

## B. Magnetic susceptibility and heat capacity

Fig. 2 (a) shows the specific heat data of the $Ca_3NiNb_2O_9$ at different field. A peak was observed around 4.2 K in the zero field curve. According to the literature on $Sr_3NiNb_2O_9$ [10] and $Ba_3NiNb_2O_9$ [21], this peak corresponded to the long range magnetic order transition. As the magnetic field increased, the peaks shifted to lower temperature and became broader. In Fig. 2(b), $C_p/T$ vs $T$ and $dC_p/dT$ vs $T$ presented two anomalies as $T_{N2}$ and $T_{N1}$, respectively, which were similar to $Sr_3NiNb_2O_9$ and should be related to the distorted lattice.

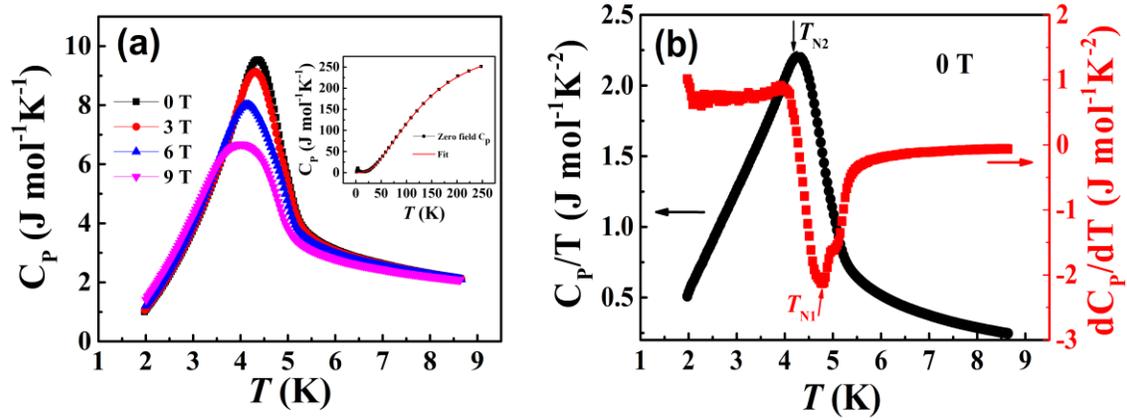

FIG. 2. (a) Temperature dependence of the specific heat for the polycrystalline $Ca_3NiNb_2O_9$ sample at different magnetic fields. The inset is the specific heat at zero field and fit - (b) Temperature dependence of $C_p/T$ and $dC_p/dT$ around $T_{N1}$ and $T_{N2}$ at zero magnetic field.

The temperature dependence DC susceptibility of $Ca_3NiNb_2O_9$ with different magnetic fields were shown in Fig. 3(a). As the field increased, the magnetic transition temperatures decreased and became broader similar to the behavior of the heat capacity anomaly. The data at 0.02 T followed the Curie-Weiss law at high temperature (inset). Fitting the $\chi(T)$ data from 100-300 K with linear Curie-Weiss law, we obtained -28 K for the Curie-Weiss temperature ($\theta_{CW}$) suggesting the dominance antiferromagnetic (AFM) exchange interactions. For $Ca_3NiNb_2O_9$, $S = 1$, the effective moment ($\mu_{eff}$) was calculated as 3.16 $\mu_B$/Ni with a corresponding Landé $g$ factor of 2.23, based on $\mu_{eff} = g\sqrt{S(S+1)}\mu_B$, which was comparable with those of $Ba_3NiNb_2O_9$, $Sr_3NiNb_2O_9$ and other compounds with $Ni^{2+}$ ions [9, 10, 21, 28, 29].

The spin frustration ratio $f$ was defined as the ratio of the absolute value of Curie-Weiss temperature ($\theta_{CW}$) to transition temperature $T_N$, thus $f = |\theta_{CW}|/T_N$. For $Sr_3NiNb_2O_9$ and $Ca_3NiNb_2O_9$ samples, we chose the higher transition temperature, $T_{N1}$, for calculation. As shown in TABLE III, the frustration ratio $f$ increased with decreasing the A-site ion radii, and $Ca_3NiNb_2O_9$ had the strongest frustration and competing interactions.

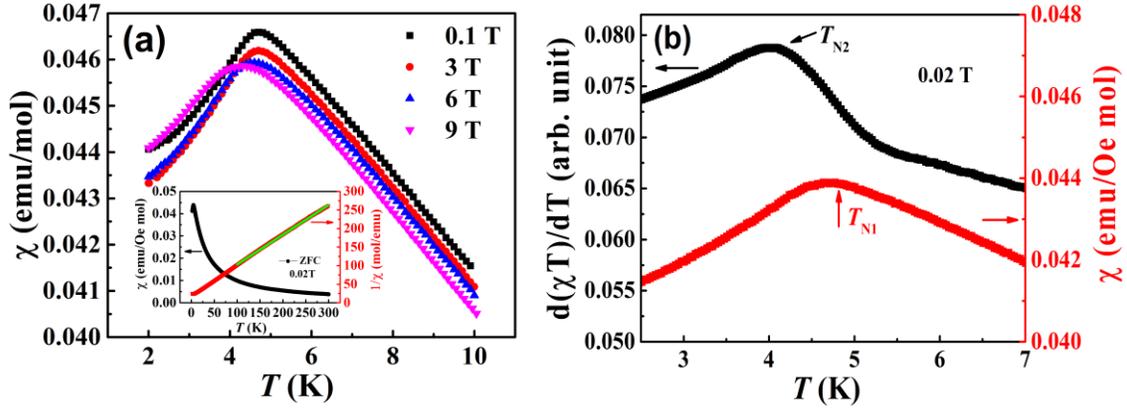

FIG. 3. (a) Temperature dependence of susceptibility around the phase transition at different fields. The inset is temperature dependence of $\chi$ and $1/\chi$ for the polycrystalline $Ca_3NiNb_2O_9$ sample at $H = 0.02$ T. The green line is the linear fit of $1/\chi$ from 100 K to 300 K. the (b) The temperature dependence of $d(\chi T)/dT$ and $\chi$ from 2.5 to 7 K.

TABLE III. The parameters extracted from the specific heat and dc susceptibility measurements for $A_3NiNb_2O_9$ ($A$ = Ba, Sr, and Ca).

| Parameters | $Ba_3NiNb_2O_9$ | $Sr_3NiNb_2O_9$ | $Ca_3NiNb_2O_9$ |
| --- | --- | --- | --- |
| $T_{N1}$ (K) | 4.9 | 5.5 | 4.8 |
| $T_{N2}$ (K) |  | 5.1 | 4.2 |
| $\theta_{CW}$ (K) | -16.4 | -21.5 | -28 |
| $\mu_{eff}$ ($\mu_B$ /Ni) | 3.15 | 3.21 | 3.16 |
| $g$ | 2.23 | 2.27 | 2.23 |
| $f$ | 3.4 | 3.9 | 5.83 |

From Fig. 3(b), the $\chi$ data exhibited a clear peak at ∼4.8 K, which was consistent with $T_{N1}$ from the $dC_p/dT$ data. Another peak could be observed from the $d(\chi T)/dT$ at ∼4.2 K, which was related to a long range magnetic ordering and corresponds to $T_{N2}$ defined from the specific heat, Fig. 2(b). It was worth noting for the Ca compound the two-step phase transition was not so strong and sensitive to different measurement techniques to some extent. For example, $T_{N2}$ transition was more sensitive to thermal measurement while $T_{N1}$ transition could be detected more easily by dc susceptibility. Fig. 4(a) compared the temperature dependence of $C_p/T$ for the polycrystalline $A_3NiNb_2O_9$ ($A$ = Ba, Sr, and Ca) samples at zero magnetic field. The Ba-compound exhibited the sharpest peak with the lambda-type feature in specific heat data, which result from its single-step phase transition at zero field. As shown in Fig. 4(b), at low magnetic fields, temperature dependence of $d(\chi T)/dT$ for the polycrystalline $Ba_3NiNb_2O_9$ sample also showed a sharp peak at $T_N$, while the two nearby $T_N$ peaks of the Sr and Ca compounds indicated a two-step transition for them.

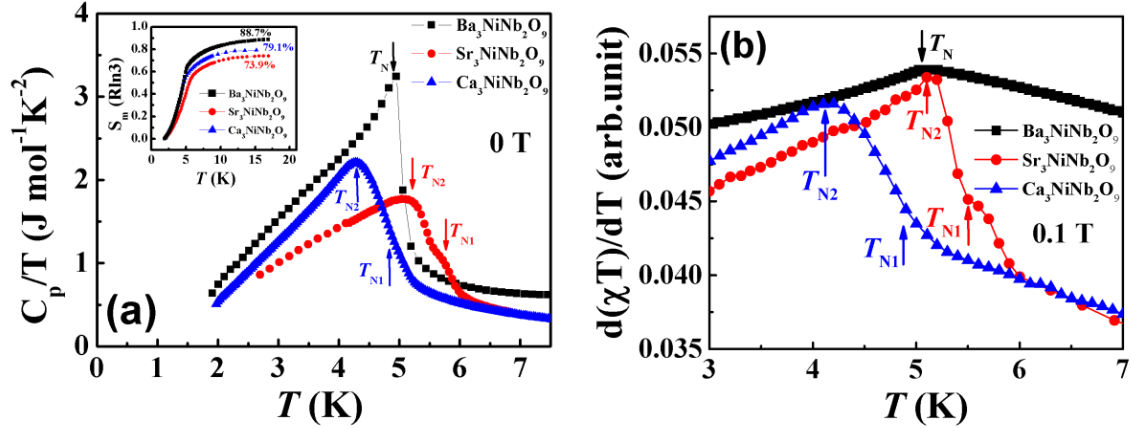

FIG. 4. (a) The comparison of the specific heat for $A_3NiNb_2O_9$ at zero magnetic field. The inset is the magnetic entropy of $A_3NiNb_2O_9$. (b) The comparison of the $d(\chi T)/dT$ for $A_3NiNb_2O_9$ at very low magnetic fields.

Based on the mean-field theory, the Heisenberg Hamiltonian, $J\sum_{\langle i,j \rangle}(S_i S_j)$, could be approximately related to the exchange $J$ and the Curie-Weiss temperature $\theta_{CW}$, $J = -3k_B\theta_{CW}/zS(S+1)$. For $A_3NiNb_2O_9$ ($A$ = Ba, Sr, and Ca), each $Ni^{2+}$ ion was surrounded by 6 NNs ($z = 6$) with the interaction $J$ in the triangular lattice, hence, $J/k_B = -3\theta_{CW}/12$, and $J_{Ba}/k_B$, $J_{Sr}/k_B$, and $J_{Ca}/k_B$ were ~ 4.1 K (0.35 meV), 5.4 K (0.46 meV), and 7.0 K (0.60 meV), respectively. Therefore, the Sr and Ca compounds would exhibit bigger $J$ than the Ba compound according to the calculated results.

The magnetic entropy of the $A_3NiNb_2O_9$ were shown in the inset of Fig. 4(a). The total magnetic entropy suggests that the degree of disorder of these three compounds were Ca >Sr >Ba, which is consistent with the behavior of the tolerance factor $t$ in TABLE I. To extract the magnetic contribution from the total heat capacity, an equation consisted of the linear combination of one Debye and several Einstein terms was used to estimate the lattice specific heat,

$$C_L(T) = C_D \left[ 9R \left(\frac{T}{\theta_D}\right)^3 \int_0^{\theta_D/T} \frac{x^4 e^x}{(e^x-1)^2} dx \right] + \sum_i C_{E_i} \left[ 3R \left(\frac{\theta_{E_i}}{T}\right)^2 \frac{\exp\left(\frac{\theta_{E_i}}{T}\right)}{\left[\exp\left(\frac{\theta_{E_i}}{T}\right) - 1\right]^2} \right]$$

where $R$ is the universal gas constant, $\theta_D$, $\theta_E$ are the Debye and Einstein temperatures, respectively. $C_D$, $C_{E_i}$ are the relative weights of the acoustic and the optical phonon contribution of the heat capacity. There are 15 atoms per formula in our system. The best fit to the data from 30 to 250 K resulted to one Debye and two Einstein terms with the proportion 1: 5: 9 for the $C_D$: $C_{E_1}$: $C_{E_2}$, and $\theta_D$=160 K, $\theta_{E_1}$=255 K, $\theta_{E_2}$=540 K. The magnetic component of the specific heat $C_m$ was obtained after subtracting the lattice contribution from the data. The magnetic entropy $S_m$ was obtained by integrating $C_m*T$ throughout the range of temperatures measured (inset of Fig. 4(a)). The $Ni^{2+}$ has a spin-1 in this compound. The total entropy saturated at about 79.1% of the Rln3. The entropy loss could be due to the quantum fluctuation in this spin frustrated system. We also measured the heat capacity of the $Sr_3NiNb_2O_9$ and $Ba_3NiNb_2O_9$, and then processed the data with the same method of $Ca_3NiNb_2O_9$. The total magnetic entropy was about 73.9% and 88.7% of Rln3 for $Sr_3NiNb_2O_9$ and $Ba_3NiNb_2O_9$, respectively. The total magnetic entropy decreased as Ba> Ca >Sr, which indicated the smallest magnetic entropy loss of Ba compound due to its smallest lattice distortion. However, for the Sr and Ca compounds, the magnetic entropy losses were dominated by the competition effect between lattice distortion and quantum fluctuation.

## C. Inelastic neutron scattering (INS)

To explore the spin dynamics of $A_3NiNb_2O_9$ ($A$ = Ba, Sr, and Ca) in details, we measured the magnetic excitations from INS spectra. Figs. 5(a)-(c) showed the powder spectrum of $A_3NiNb_2O_9$ ($A$ = Ba, Sr, and Ca) at 1.5 K, below $T_N$. Similar as the S-1/2 TLAF compound, $Ba_3CoSb_2O_9$, both gapped and gapless modes were observed in the magnetic DOS [27]; meanwhile, unlike $Ba_3CoSb_2O_9$, there was no obvious continuum observed at higher energy, which might be due to the larger spin moment ($S = 1$) and the weak signal from powder average effect.

The low-energy magnon bandwidth of $Ba_3NiNb_2O_9$ was around 1.0 meV, which was lower than those of $Sr_3NiNb_2O_9$ and $Ca_3NiNb_2O_9$ (~1.25 meV). The magnetic DOS of the $A_3NiNb_2O_9$ ($A$ = Ba, Sr, and Ca) system showed minimums around $Q \approx 0.85$ Å$^{-1}$, which notably corresponded to commensurate magnetic wave vector $\mathbf{q_m}$ = [1/3, 1/3, 1/2]. Moreover, the gaps of this branch at $\mathbf{q_m}$ = [1/3, 1/3, 1/2] for $Ba_3NiNb_2O_9$ and $Ca_3NiNb_2O_9$ was around 0.6 meV, while $Sr_3NiNb_2O_9$ exhibited a similar gap of around 0.8 meV, Fig.6. The magnetic signals were momentum dependent due to stronger ridges of intensity at $Q \approx 0.85$ Å$^{-1}$ than $Q \approx 1.5$ Å$^{-1}$ and $Q \approx 2.0$ Å$^{-1}$.

The INS results were simulated by the linear spin-wave (LSW) theory based on the quasi-2D XXZ Hamiltonian on a vertically stacked triangular lattice, Figs. 5(d)-(f). The appropriate Heisenberg Hamiltonian was,

$$H = J \sum_{\langle i,j \rangle}^{\text{layer}} \left( S_i^x S_j^x + S_i^y S_j^y + \Delta S_i^z S_j^z \right) + J' \sum_{\langle l,m \rangle}^{\text{interlayer}} \mathbf{S}_l \cdot \mathbf{S}_m$$

where $J$ and $J'$ ($J, J' > 0$) were the intra- and inter-layer NN antiferromagnetic exchange energies, respectively. $\Delta$ ($0 < \Delta \leq 1$) was the easy-plane exchange anisotropy from the same 120° structure as in the Heisenberg case. TABLE IV was the summary of the parameters in the Hamiltonian according to the LSW approximation.

The E-integrated scans were obtained in Fig. 6, INS measurements and powder averaged LSW approximation have a good agreement: the data of Ca compound could be explained as the most Heisenberg-like feature and the best fit was obtained.

In the $A_3NiNb_2O_9$ system, the Sr compound exhibited the largest intra-layer NN exchange ($J = 0.36$ meV), while the Ca compound obtained the largest inter-layer exchange

($J' = 0.3J$). Hence, the intra- and inter-layer NN exchange satisfied these relations, $J_{Ba} < J_{Ca} < J_{Sr}$ and $J'_{Ba} < J'_{Sr} < J'_{Ca}$, respectively. And the mean-field-theory doesn't work for $Sr_3NiNb_2O_9$ and $Ca_3NiNb_2O_9$, as shown in TABLE IV. The biggest gap for the Sr compound (~0.8 meV) suggested the largest intra-plane NN exchange anisotropy than those of ~0.6 meV for the Ba and Ca compounds. When the $Ba^{2+}$ ion was substituted with the smaller $Sr^{2+}$ ion, the $Ni^{2+}$ triangle changed from an equilateral triangle to an isosceles triangle, as shown in Fig. 1(e). The smaller A-site ions resulted in smaller triangle lattice and larger intra-plane exchange anisotropy. When the smallest $Ca^{2+}$ ion was on the A-site, the lattice distortion was enhanced further, and the inter-plane interaction was strong enough to affect the intra-plane anisotropy, which presented the weak quantum effect. From the NPD data, the distances between the triangular *ab*-planes for the Ba, Sr and Ca compounds were 7.034 Å, 6.924 Å and 6.815 Å, respectively. And the ratios of the average intra-plane Ni-Ni bond lengths to these inter-layer distances were ~0.81 for the Ba and Sr compounds, and ~0.80 for the Ca one. Therefore, the shortest inter-layer distances and the smallest ratios of bond lengths to plane distances for the Ca compound led to very strong inter-plane coupling with the largest $J'/J = 0.30$ and reduced the $J$ slightly to 0.31 meV (< $J_{Sr} = 0.36$ meV). In addition, the Ca compound showed more obvious gapless feature than the Sr compound from the fitting. The best fitting parameters of easy-plane exchange anisotropy were $\Delta = 0.95$ and $\Delta = 0.75$ for $Ca_3NiNb_2O_9$ and $Sr_3NiNb_2O_9$, respectively, as shown in TABLE IV. It indicated that the Ca-compound had less easy plane anisotropy compared to the Sr one due to its more dispersive feature with the similar bandwidth as the Sr compound.

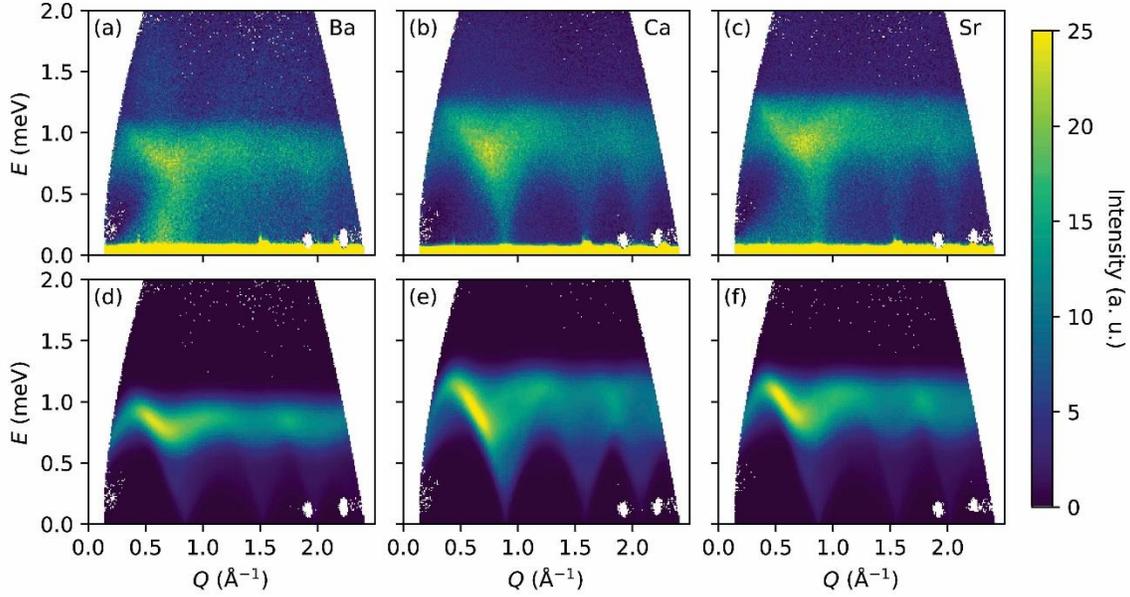

FIG. 5. Powder spectra measured experimentally at 1.5 K on spectrometer NEAT: (a) $Ba_3NiNb_2O_9$, (b) $Ca_3NiNb_2O_9$, (c) $Sr_3NiNb_2O_9$ powder spectra at 1.5 K. The powder-average LSW approximation of (d) $Ba_3NiNb_2O_9$, (e) $Ca_3NiNb_2O_9$, (f) $Sr_3NiNb_2O_9$.

The competition of the easy-plane exchange anisotropy $\Delta$ and $J'/J$ was treated as the dominant effect in the $A_3NiNb_2O_9$ system. Moreover, we also considered the easy-axis anisotropy, single ion anisotropy and DM effects in the system by the LSW approximation: 1) Easy-axis anisotropy could lift the entire band up, create a gap and modify the ordering wave vector; 2) If single ion anisotropy existed, the out of plane canting would be bigger and towards the collinear structure; 3) the DM interaction could open a gap at the wave vector for a single-crystal spectrum with the gapless (Goldstone) mode persisting at the $\Gamma$-point. Since the Hamiltonian still had continuous symmetry in the *ab* plane, and the ground state could break it spontaneously. No matter how weak the powder spectra was, it would always be gapless.

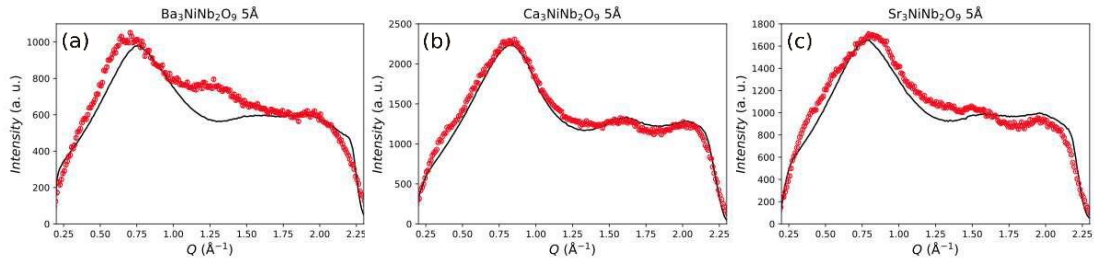

FIG. 6. Comparison between experiment (red dots) and LSWT (the black line) simulation shown as energy-integrated within $0.2 \leq E \leq 1.5$ meV for: (a) $Ba_3NiNb_2O_9$, (b) $Ca_3NiNb_2O_9$, (c) $Sr_3NiNb_2O_9$.

TABLE IV. The summary of the exchange parameters in the Hamiltonian according to the LSW approximation.

| Parameters | $Ba_3NiNb_2O_9$ | $Ca_3NiNb_2O_9$ | $Sr_3NiNb_2O_9$ |
| --- | --- | --- | --- |
| $J^{MFT}$ (meV) | 0.35 | 0.60 | 0.46 |
| $J$ (meV) | 0.30 | 0.31 | 0.36 |
| $J'$ (meV) | 0.015 | 0.093 | 0.018 |
| $J'/J$ | 0.05 | 0.30 | 0.05 |
| $\Delta$ | 0.70 | 0.95 | 0.75 |

## IV. DISCUSSION

To understand the intralayer antiferromagnetic interaction of the $Ni^{2+}$ ions, we examined the superexchange interaction in the structure using Goodenough-Kanamori's theoretical framework [30], which discussed the relation between the symmetry of electron orbitals and superexchange interaction via a nonmagnetic anion. In $Ba_3NiNb_2O_9$, two superexchange pathways for the $Ni^{2+}$ spins in the same layer were shown in Fig. 7(a), $Ni^{2+}$-$O^{2-}$-$O^{2-}$-$Ni^{2+}$ and $Ni^{2+}$-$O^{2-}$-$Nb^{5+}$-$O^{2-}$-$Ni^{2+}$, respectively, by sharing the corner oxygens of the $NiO_6$ and $NbO_6$ octahedrons. Although the first $Ni^{2+}$-$O^{2-}$-$O^{2-}$-$Ni^{2+}$ superexchange pathway could be observed as the antiferromagnet, the connection of $O^{2-}$-$O^{2-}$ need to hybridize/distort the *s*-orbital or the other *p*-orbitals of $O^{2-}$. For the second superexchange path of $Ni^{2+}$-$O^{2-}$-$Nb^{5+}$-$O^{2-}$-$Ni^{2+}$, Fig. 7(b), the superexchange interaction between the spins on the $d_{x^2-y^2}$ orbitals of the $Ni^{2+}$ ions were considered. From the Rietveld refinements of neutron powder diffraction, the $O^{2-}$-$Nb^{5+}$-$O^{2-}$ bond angle is very close to 90° at 91.13°, and the $Ni^{2+}$-$O^{2-}$-$Nb^{5+}$ is 180°. In this case, the spin 1 on the left $Ni^{2+}$ ion transferred to the $2p_y$ orbital of the $O^{2-}$ by combining with the $2p_y$ orbital of the filled outermost $Nb^{5+}$ 4*p* orbitals to form a molecular orbital while the spin 2 on the right $Ni^{2+}$ ion transferred to the molecular orbital composed of the $2p_x$ orbital of the $O^{2-}$ and the $Nb^{5+}$ ions. According to Hund's rule, these two spins on the $p_y$ and $p_x$ orbitals must be parallel. Thus, a ferromagnetic superexchange interaction formed between these two $Ni^{2+}$ ions after these two spins transferred back to the $d_{x^2-y^2}$ orbitals of $Ni^{2+}$. Therefore, the $Ni^{2+}$-$O^{2-}$-$O^{2-}$-$Ni^{2+}$ exchange had a shorter connection path and thereby a stronger orbital-distortion than the one of $Ni^{2+}$-$O^{2-}$-$Nb^{5+}$-$O^{2-}$-$Ni^{2+}$. For $Ba_3NiNb_2O_9$, the distortion effect was not dominant, and the AFM interaction overcame the FM interaction. As the magnetic field was applied, the distortion effect was strengthened, and the FM interaction could be observed in the *uud* phase [10]. Similar competitive superexchange interactions in other TLMs with layered perovskite structures have been reported. For example, in $Ba_3CoNb_2O_9$, the AFM $Co^{2+}$-$O^{2-}$-$O^{2-}$-$Co^{2+}$ interaction is stronger than the FM $Co^{2+}$-$O^{2-}$-$Nb^{5+}$-$O^{2-}$-$Co^{2+}$ interaction, which resulted in a weak AFM interaction [31]. However, in other triangular magnet systems, such as, $AAg_2M(VO_4)_2$ ($A$ = Ba, Sr; $M$ = Co, Ni) [32], the AFM $Co^{2+}$-$O^{2-}$-$O^{2-}$-$Co^{2+}$ interaction was weaker than the FM $Co^{2+}$-$O^{2-}$-$V^{5+}$-$O^{2-}$-$Co^{2+}$ interaction, and $YCr(BO_3)_2$, the AFM $Cr^{3+}$-$O^{2-}$-$O^{2-}$-$Cr^{3+}$ interaction was also weaker than the FM $Cr^{3+}$-$O^{2-}$-$Yb^{3+}$-$O^{2-}$-$Cr^{3+}$ interaction [33].

As shown in Figs. 7(c) and (d), in $Sr_3NiNb_2O_9$ and $Ca_3NiNb_2O_9$, the lattice distortion resulted to the variation of the $Ni^{2+}$-$O^{2-}$ bond lengths. Most of the $O^{2-}$-$Nb^{5+}$-$O^{2-}$ bond angles were distorted away from the ideal value of 90°, to as small as ~76° and some as large as ~104°. The $Ni^{2+}$, $O^{2-}$ and $Nb^{5+}$ ions were not on the same line. Thus, the FM interactions were influenced by the distortion and led to stronger resultant AFM interactions. To obtain the *uud* phase, a higher magnetic field needed to be applied [21]. Meanwhile, the polarization of Sr and Ca compound should be smaller than that of Ba compound, and the re-entrant signal in pyroelectric current of Ba compound was absent [21].

Moreover, only one transition was observed in $Ba_3NiNb_2O_9$, which was consistent with the easy-plane anisotropy in this isotropic system. For $Sr_3NiNb_2O_9$ and $Ca_3NiNb_2O_9$ with distorted structures, the easy-axis anisotropy was excluded by the LSW approximation of the INS measurements while two magnetic transitions were observed in both compounds. An extra noncollinear magnetic phase existed at low temperature, which corresponded to the phase transition at $T_{N1}$, and the spin structure of this phase should be 120° in *ab* plane. For the $S$ = 1 case, the high temperature transition from the paramagnet was not clear and a different phase diagram had been proposed by theory, specifically the stripe phase related to a spin reorientation in honeycomb lattice [34], and the zigzagging stripe phase in the isosceles triangular networks by the harmonic particle interaction [35].

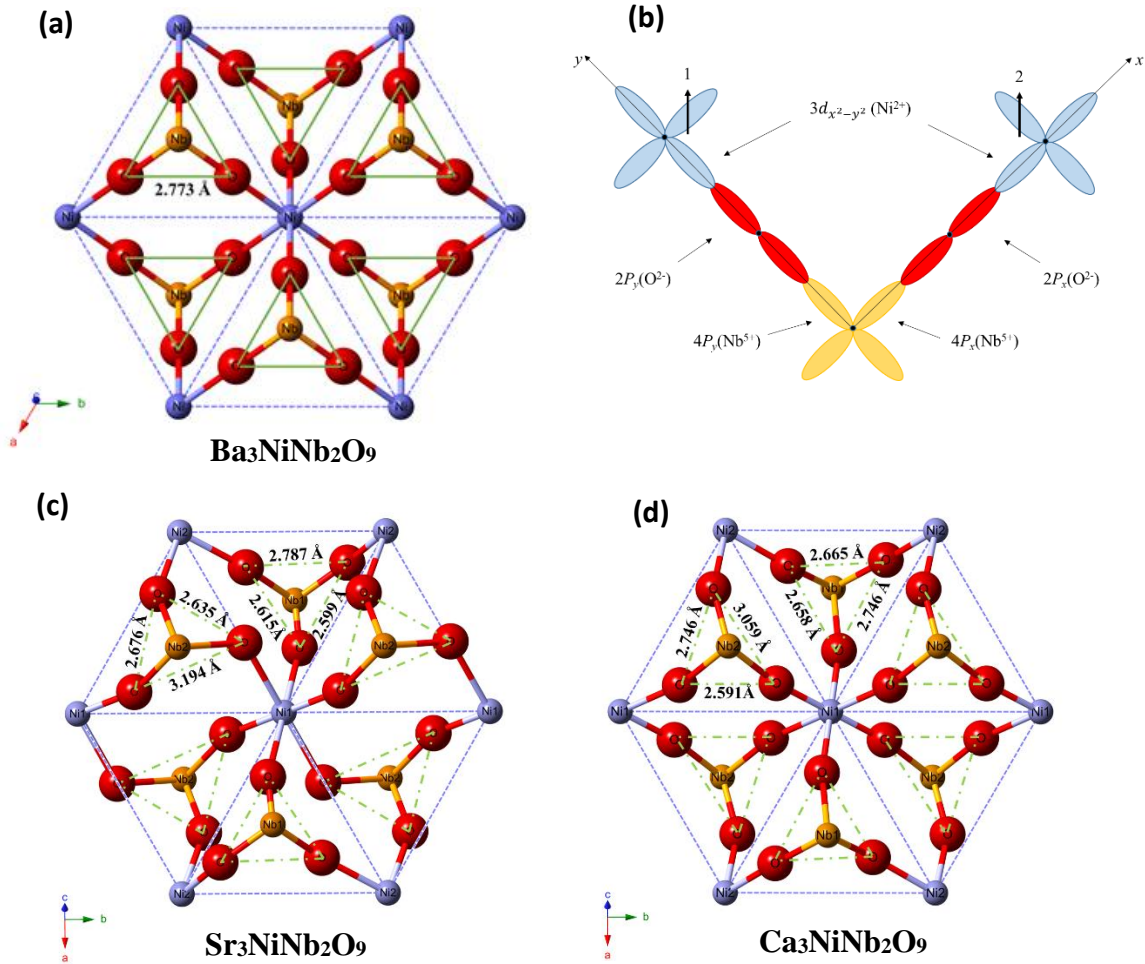

FIG. 7. (a) Superexchange pathways for FM $Ni^{2+}$-$O^{2-}$-$O^{2-}$-$Ni^{2+}$ and AFM $Ni^{2+}$-$O^{2-}$-$Nb^{5+}$-$O^{2-}$-$Ni^{2+}$ interactions in $Ba_3NiNb_2O_9$. (b) Orbital configurations of FM interactions through $Ni^{2+}$-$O^{2-}$-$Nb^{5+}$-$O^{2-}$-$Ni^{2+}$ superexchange pathway in $Ba_3NiNb_2O_9$. (c) Superexchange pathways in $Sr_3NiNb_2O_9$. (d) Superexchange pathways in $Ca_3NiNb_2O_9$.

## V.    CONCLUSIONS

In summary, the lattice distortion effect on the magnetic ground states of spin-1 TLAF was investigated by comparing two isosceles triangular lattice antiferromagnets $Sr_3NiNb_2O_9$ and $Ca_3NiNb_2O_9$ with an equilateral triangular compound $Ba_3NiNb_2O_9$. Although the effective magnetic moment across the family is the same, the magnetic frustration of the system increases from the equilateral triangle to the isosceles one. Moreover, the two magnetic phase transitions were observed in $A$ = Sr and Ca compared to one in $A$ = Ba. However, the lattice distortion did not tune the easy plane anisotropy of the Ba compound. Instead, the lattice distortion generated an extra competitive magnetic phase with the 120° spin structure that has an out of *ab* plane canting at low temperatures for both Sr and Ca compounds and yielded larger inter-plane exchange energy with bigger anisotropy for the ground state of the spin-1 $A_3NiNb_2O_9$ TLAF.

**Acknowledgement**: J.M. and G.H.W. acknowledge support from the NSF China (grant 11774223). Thank the support from NSF-DMR-1350002. Research conducted at ORNL's High Flux Isotope Reactor was sponsored by the Scientific User Facilities Division, Office of Basic Energy Sciences, United States Department of Energy.